\documentclass{article}

\usepackage[preprint,nonatbib]{nips_2018}

\usepackage[utf8]{inputenc} 
\usepackage[T1]{fontenc}    
\usepackage{hyperref}       
\usepackage{booktabs}       
\usepackage{amsfonts}       
\usepackage{nicefrac}       
\usepackage{microtype}      

\usepackage{multicol}
\usepackage{multirow}
\usepackage{graphicx}

\usepackage{xcolor}
\usepackage{soul}

\title{Document classification using a Bi-LSTM to unclog Brazil's supreme court}

\author{
  F.A.\ Braz, N.C.\ Silva, T.E.\ de Campos\thanks{The authors belong to three diferent insitutions of UnB: Faculdade de Engenharias do Gama, Departamento de Ci\^encia da Computa\c{c}\~ao and Faculdade de Direito. Further information at https://cic.unb.br/\~{}teodecampos/ViP}, F.B.S.\ Chaves, M.H.P.\ Ferreira,\\
  {\bf P.H.G.\ Inazawa,  V.H.D.\ Coelho, B.P.\ Sukiennik, A.P.G.S. Almeida, F.B.\ Vidal, D. Alves Bezerra,}\\
  {\bf D.B.\ Gusm\~ao, G.G.\ Ziegler, R.V.C.\ Fernandes, R.\ Zumblick, F.\ Hartmann Peixoto} \\
  Universidade de Bras\'{\i}lia (UnB), DF, Brazil\\
  \texttt{\{fabraz,niltoncs,teodecampos\}@unb.br}\\
}

\begin{document}

\maketitle

\begin{abstract}
  The Brazilian court system is currently the most clogged up judiciary system
  in the world. Thousands of lawsuit cases reach the supreme court every day.
  These cases need to be analyzed in order to be associated to relevant
  tags and allocated to the right team.
  Most of the cases reach the court as raster scanned documents with widely 
  variable levels of quality. 
  One of the first steps for the analysis is to classify these documents.
  In this paper we present a Bidirectional Long Short-Term Memory network (Bi-LSTM) 
  to classify these pieces of legal document. 
\end{abstract}

\section{Introduction}
Ensuring equal access to justice for all is a real challenge in Brazil due to the sluggishness of judicial process and the high volume of new cases entering in the justice system every year. 
In the Brazilian higher courts, the average time to reach the sentence is 11 months in the Superior Court of Justice (STJ), 1 year and 2 months in the Superior Court of Labor (TST) and 8 months in the Superior Electoral Court (TSE). This scenario is worsening each year: around 28.8 million new cases reach the Brazilian Judiciary and there are already approximately 79.6 million other cases in stock in Brazilian courts, related to 2016. Adding up these figures, there are
approximately 108.4 million legal processes being carried on by Brazilian judiciary~\cite{lucia_JusticaEmNumeros_2017}.
This overload has a cost of USD 22 billion per year, according to National Justice Council (CNJ), and is one of the main causes of legal insecurity and criminal impunity in the country. The congestion rate was 73.0\% in 2016. That means that only 27\% of all the cases processed were solved in that year. 

Our work aims to contribute towards goal 16.3 of the UN agenda 2030: 
``ensure equal access to justice for all'' \cite{un_Agenda2030_2015}. 
To overcome the obstacles that prevent the Brazilian judicial system from democratically guaranteeing the right 
to due process of law, we propose to apply machine learning methods 
to classify legal processes that reach
the Brazilian Supreme Court (in Portuguese, \textit{Supremo Tribunal Federal -
STF})~\cite{stf_press_30_05_2018,stf_press_30_08_2018}.
According to the STF, it would take 22,000 man-hours by its
employees and trainees to analyze the approximately 42,000 processes
received per semester~\cite{lucia_JusticaEmNumeros_2017}. 
The court also points out that the time its
employees spend on classifying these processes could be better applied
at more complex stages of the judicial work flow.

Most of the cases reach the court are in the form of PDF files with 
raster scanned documents. About 10\% of them are completely unstructured 
volumes which encloses several documents per file without any digital index. 
In Brazil's judiciary, documents that compose a legal case are grouped into
briefs or parts (called {\em pe\c{c}as} in Portuguese) which are categorized into a set of classes. 
In order to speed up the analysis of cases, the first step needed is to automate the classification
of these documents. 

This paper reports results of a preliminary evaluation on a dataset containing
6,814 documents ({\em pe\c{c}as}) from the STF.
We propose a Bidirectional Long Short-Term Memory network architecture for this task
and show that it obtains 84\% F$_1$ score on this dataset with no pre-processing.

\section{Proposed method \label{sec:methodology}}
\subsection{Text extraction}

The first step is to extract text from the PDF file.
If the content of a document page is a raster image, 
we apply the Tesseract OCR system \cite{smith_Tesseract_icdar2007}
and store the text. 
If the page embeds its text as metadata, its quality is verified using regular expressions. If the quality level is acceptable the text is stored, otherwise the OCR is applied as if the page was in raster image format and the result is stored. 


\subsection{Bidirectional Long Short-Term Memory Model}
 \label{sec:bilstm}
 The proposed model is based on recurrent models to deal with text as sequential information. Long Short-Term Memory (LSTM) models use the information from the previous status of neurons \cite{hochreiter_schmidhuber_origin_lstm_nc_1997}. More contextual information can be extracted using a bidirectional LSTM (Bi-LSTM) \cite{graves_framewise_origin_bilstm_nn_2005}. 
 The data processing flows forward and backward at same time  \cite{schuster_etal_bidirectional_TSP_1997} and the output of each LSTM are merged using their sum \cite{graves_framewise_origin_bilstm_nn_2005}.

Bi-LSTM models have shown to be effective in 
problems of speech recognition \cite{Ghaeini_etal_lstm_arXiv_2018, Kiperwasser_etal_lstm_arXiv_2016,Tai_etal_lstm_arXiv_2015,rao_spasojevic_lstm_arxiv,huang_etal_lstm_arXiv_2015},
being a state-of-the-art model to classify sequential data into multiple classes. Documents are sequences of words and our problem is also multi-class, therefore Bi-LSTM is a suitable model. 

\label{sec:architecture}

\begin{figure}[htb]
  \centerline{\includegraphics[width=.7\columnwidth]{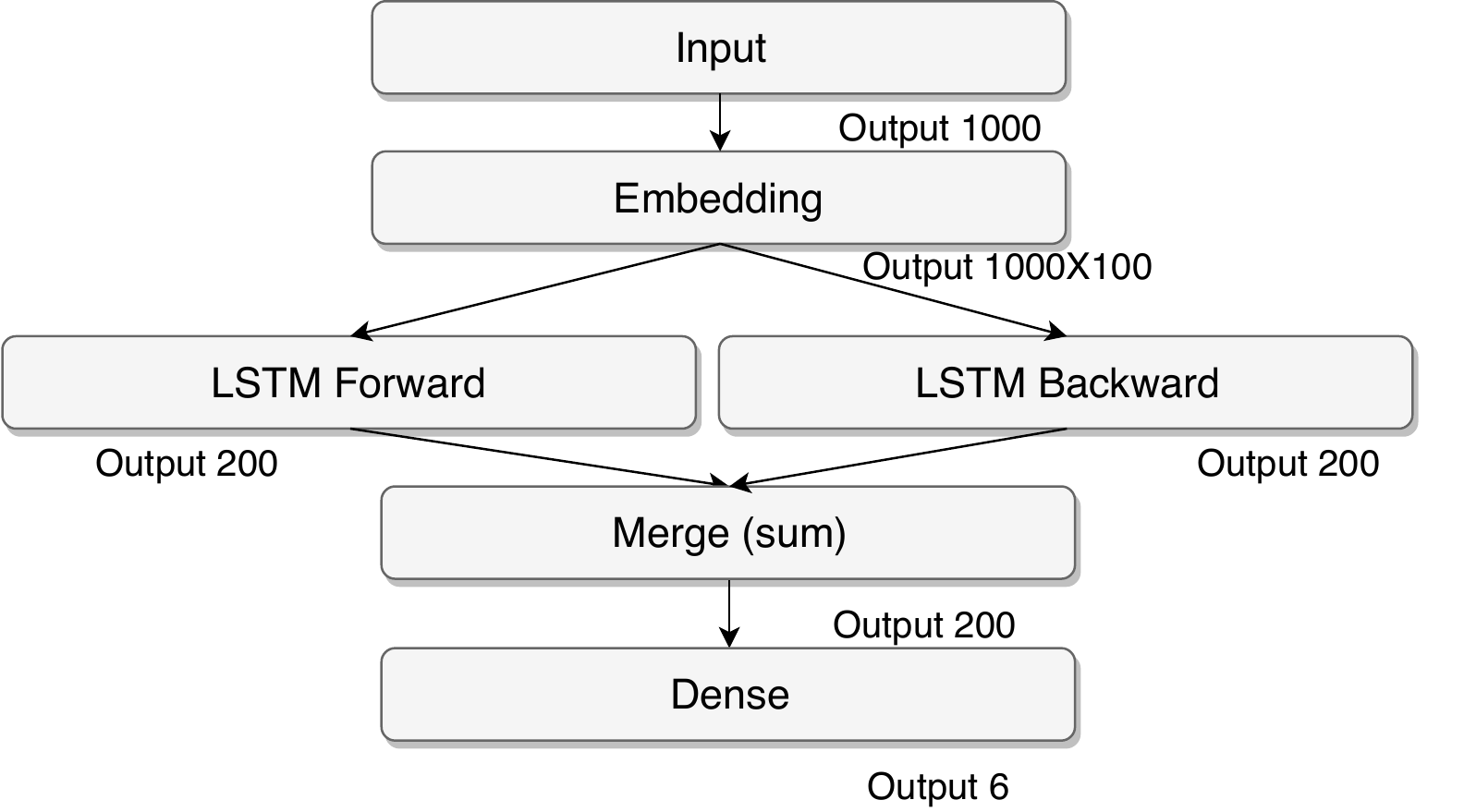}}
  \caption{Bi-LSTM Architecture diagram}
  \label{fig:diagram}
\end{figure}

Our architecture (shown in Figure~\ref{fig:diagram}) is composed of three main layers with 1000 tokens of input. 
We followed \cite{rao_spasojevic_lstm_arxiv,huang_etal_lstm_arXiv_2015} and used 
word embedding as an input layer of the Bi-LSTM. This layer 
transforms each token in a distributed array of 100 dimensions.
The recurrent layer has two hidden LSTM, a forward and backward layer model,
each with 200 memory blocks and one-cell. The output of this layer uses a ReLu activation. The two hidden LSTMs are combined by adding their outputs. 
The last layer is dense, with 6 output neurons and a Softmax activation function.

\section{VICTOR project's dataset of brief parts ({\em pe\c{c}as})}
\label{sec:dataset}

We used the dataset described in \cite{daSilva_Braz_CNN_ICoFS_2018}.
Our work focuses on classifying five main types of legal documents handled by the Brazilian supreme court (STF). Documents that do not belong to these classes are grouped in a class called {\em Others}. Table \ref{tab:dataset} presents the classes used and the number of samples in each of them.

\begin{table}[htb]
    \caption{Set of classes in VICTOR project's dataset of document types ({\em pe\c{c}as}). }
    \label{tab:dataset}
    \centering
    \begin{tabular}{rlr}
    \hline 
    Label (in Portuguese) & Description & Samples \\
    \hline
{\em Acórdão} & Appelate Decision & 82                                    \\
{\em Recurso Extraordinário (RE)} & Extraordinary Appeal & 63             \\
{\em Agravo de Recurso Extraordinário (ARE)} & Extraordinary Appeal Bill/Review & 92 \\
{\em Despacho} & Administrative Orders & 55                               \\
{\em Sentença} & Judgement & 110                                          \\
{\em Outros} & Other Documents Types & 280                                \\
\hline
    \end{tabular}
\end{table}


A total of 6,814 text documents were manually labeled by a team of four specialist lawyers. 
This dataset was split into three parts: 70\% of the samples for training, 20\% for validation and 10\% for test. 


This dataset is quite challenging as it has a high level of within class diversity. 
The documents do not follow a standard, not only in terms of layout but also in terms of
scanned image quality and whether or not they embed digital text or have only raster images.
Furthermore, a significant part of these documents often contained handwritten annotations, stamps, stains etc.
As discussed earlier, the PDF files are often not indexed and some of them include several documents.

\section{Experiments and results}
Although legal documents can be quite long, the first page is usually 
the most informative. Furthermore, the files in VICTOR dataset often 
include multiple documents, so later pages in a PDF file can be
unreliable. For these reasons, our Bi-LSTM model was designed to take 
inputs of 1000 tokens, which usually covers most of the contents of one page.
Our evaluations on the validation set show that this was discriminative enough.
One major advantage of this w.r.t.\ using text from the whole document is that
the main bottleneck of our system is the OCR system, which takes 1s per processing 
core to run on each page. By using a method that efficiently exploits the
first 1000 tokens, we only need to run the OCR on up to two pages per document.



Since most of the text was obtained by running an OCR raster scanned noisy documents (rather than extracting text from PDFs metadata),
many out-of-vocabulary `words' appeared in the text. One can deal with this problem by
using a number of regular expressions and stemming, as done in \cite{daSilva_Braz_CNN_ICoFS_2018}
and/or by running a named entity recognition system such as that of \cite{luz_etal_propor2018}. 
In this paper, we simply limited the tokenizer's vocabulary to the 100,000 most frequent distinct words.
This was enough to include all relevant words as well as specific symbols of the judiciary system,
such as law numbers and Latin words.


Our model was trained for 20 epochs with a batch size of 64 samples and a learning rate of 0.001. The total training time was of 120.02s on a NVidia Titan XP, which has 12GB of RAM.
In this setup, prediction on is done in 1.47ms per document. In contrast,
the CNN model of \cite{daSilva_Braz_CNN_ICoFS_2018} takes 5.87ms per document,
excluding the time to run the OCR (1s per page per CPU core). 



\begin{figure}[htb]
\centering
\begin{minipage}[b]{.4\textwidth}
\begin{tabular}{l|lll}
    \hline  
                    & prec. & rec. & F$_1$ \\ 
        \hline
        ARE         & 0.82      & 0.84   & 0.83  \\ 
        Acordão     & 0.71      & 0.89   & 0.79  \\ 
        Desp.       & 0.74      & 0.82   & 0.78 \\ 
        Outro       & 0.91      & 0.82   & 0.87 \\ 
        RE          & 0.77      & 0.70   & 0.73 \\ 
        Sent.       & 0.92      & 0.95   & 0.93 \\ 
        \hline
        average     & 0.85      & 0.84   & 0.84  \\ 
        \hline
\end{tabular}
\caption{\label{tab:results}Bi-LSTM results per class.}
\end{minipage}
\begin{minipage}[b]{.575\textwidth}
\includegraphics[width=\textwidth]{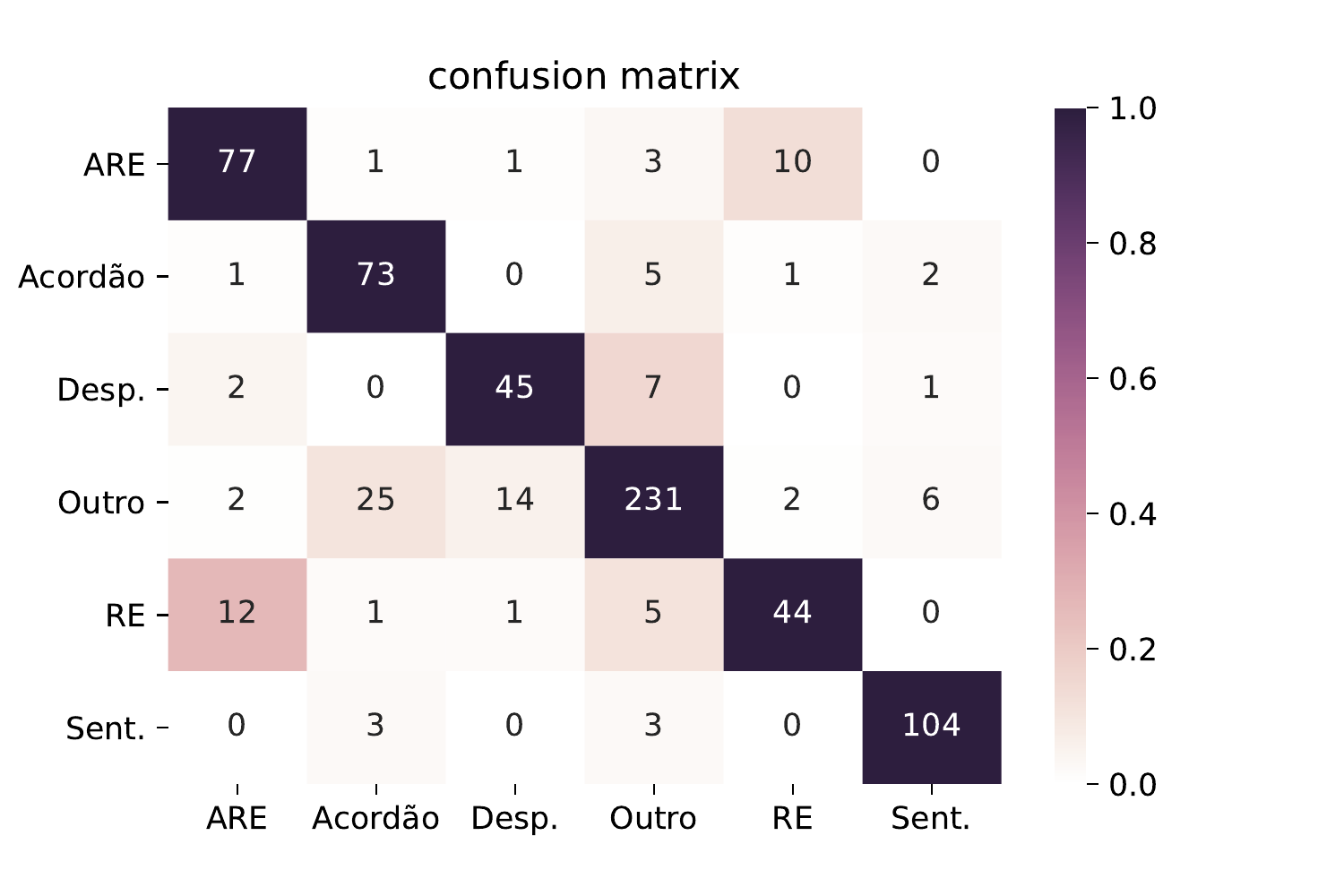}
\caption{Confusion matrix.\label{fig:matrix}}
\end{minipage}
\end{figure}

Figures \ref{tab:results} and \ref{fig:matrix} detail our results. 
Our BI-LSTM model archives a mean precision of 85\% and f$_1$-Score 84\%
without requiring any preprocessing heuristics and regular expressions. 
This contrasts with the CNN model of \cite{daSilva_Braz_CNN_ICoFS_2018},
which uses a set of hand crafted rules in the tokenization process as well
as regular expressions to remove noisy text.





\section{Conclusion}
\label{sec:conclusion}
 
 We proposed a tool to significantly speed up the first steps of the
 analysis of legal documents that reach the {\em Supremo Tribunal Federal} (STF, from Brazil), which is the world's most clogged up supreme court.
The task consists in classifying legal briefs ({\em pe\c{c}as}) into a set of 6 classes. 
For that we introduced a Bidirectional Long Short-Term Memory model 
which processes the first 1000 tokens of the documents, i.e., usually 
just the first page. This model is strong enough to classify these documents
with an F$_1$ score of 84\%, dismissing the need to run an OCR on the remaining
pages of the document.


The next step of this project consists in designing a 
tool that combines information from all documents that compose a legal 
case in order to aid the decision making in the judgment process.

\subsubsection*{Acknowledgments}
We acknowledge the support of ``Projeto de Pesquisa \& Desenvolvimento de  aprendizado de máquina (machine learning) sobre dados judiciais das repercussões gerais do Supremo Tribunal Federal - STF''. 

\medskip

\small
\bibliographystyle{plain}

\end{document}